\documentclass[twocolumn]{aastex631}
\usepackage{hyperref}
\usepackage{multirow}
\usepackage{longtable}
\usepackage{amsmath}
\usepackage{graphicx}
\usepackage[flushleft]{threeparttable}
\usepackage{amssymb}
\usepackage{amsfonts}
\usepackage{bm}

\newcommand{\bs}{\boldsymbol}

\shorttitle{Theoretical systematics in the quasar dipole}
\shortauthors{Guandalin, Piat, Clarkson \& Maartens}
\graphicspath{{./}{figures/}}

\begin{document}

    \defcitealias{boyle2000}{Boyle et al. 2000}
    \defcitealias{ross2013}{Ross et al. 2013}
    \defcitealias{croom2009}{Croom et al. 2009}
    \defcitealias{croom2004}{Croom et al. 2004}
    \defcitealias{palanque2016}{PD2016}
    \defcitealias{palanque2011}{PD2011}

    \title{Theoretical systematics in testing the Cosmological Principle with the kinematic quasar dipole}
     
    \correspondingauthor{Caroline Guandalin}
    \email{c.m.guandalin@qmul.ac.uk}
    
    \author[0000-0003-1490-9314]{Caroline Guandalin}
    \affiliation{School of Physical \& Chemical Sciences, Queen Mary University of London,  London E1 4NS, UK}
    
    \author{Jade Piat}
    \affiliation{Aix-Marseille University, Marseille, France}
    \affiliation{School of Physical \& Chemical Sciences, Queen Mary University of London,  London E1 4NS, UK}
    
    \author[0000-0001-7363-0722]{Chris Clarkson}
    \affiliation{School of Physical \& Chemical Sciences, Queen Mary University of London, London E1 4NS, UK}
    \affiliation{Department of Physics \& Astronomy, University of Western Cape, Cape Town 7535, South Africa}
    \affiliation{Department of Mathematics and Applied Mathematics, University of Cape Town 7701, South Africa}
    
    \author[0000-0001-9050-5894]{Roy Maartens}
    \affiliation{Department of Physics \& Astronomy, University of Western Cape, Cape Town 7535, South Africa}
    \affiliation{Institute of Cosmology \& Gravitation, University of Portsmouth, Portsmouth PO1 3FX, UK}
    \affiliation{National Institute for Theoretical \& Computational Sciences (NITheCS), Cape Town 7535, South Africa}



\begin{abstract}
    The Cosmological Principle is part of the foundation that underpins the standard model of the Universe. In the era of precision cosmology, when stress tests of the standard model are uncovering various tensions and possible anomalies, it is critical to check the viability of this principle. A key test is the consistency between the kinematic dipoles of the cosmic microwave background and of the large-scale matter distribution. Results using radio continuum and quasar samples indicate a rough agreement in the directions of the two dipoles, but a larger than expected amplitude of the matter dipole. The resulting tension with the radiation dipole has been estimated at $\sim 5\sigma$ for some cases, suggesting a potential new cosmological tension and a possible violation of the CP. However, the standard formalism for predicting the dipole in the two-dimensional projection of sources overlooks possible evolution effects in the luminosity function. In fact, radial information from the luminosity function is necessary for a correct projection of the three-dimensional source distribution. Using a variety of current models of the quasar luminosity function, we show that neglecting redshift evolution can significantly overestimate the relative velocity amplitude.
    While the models we investigate are consistent with each other and with current data, the dipole derived from these, which depends on derivatives of the luminosity function, can disagree by more than $3\sigma$. This theoretical systematic bias needs to be resolved before robust conclusions can be made about a new cosmic tension.
\end{abstract}



\section{Introduction}

    The Cosmological Principle (CP), i.e. that the spatial distribution of matter and radiation in the Universe is statistically homogeneous and isotropic on large enough scales, is perhaps the most fundamental assumption in modern cosmology. It is the basis for modelling the Universe using the Friedmann-Lemaître-Robertson-Walker (FLRW) metric \citep[see, e.g.,][]{ehlers1968,ellis1983,stoeger1995,Clarkson:2010uz,Clarkson:2012bg}. The CP implies that there is a unique cosmic frame defined by observers who measure statistical isotropy and homogeneity of the cosmic microwave background (CMB) and of the  matter distribution on sufficiently large scales. 
    
    Earth observers are not at rest in the cosmic frame, but are moving with a velocity $\bs{v}_o^{\rm CMB}$ relative to the CMB rest frame, i.e., the frame in which the CMB dipole vanishes and the CMB is statistically isotropic \citep{stewart1967,peebles1968}. This leads to a dipole of $3362.08 \pm 0.99\,\,\mu\,$K,  much larger than the $\ell\geq2$ multipoles, which is used to extract a velocity of $v_o^{\rm CMB}=369.82 \pm 0.11\,{\rm km/s}$ towards $(l, b) = (264.021 \pm 0.011, 48.253 \pm 0.005)^{\circ}$ \citep{aghanim2020}.
    
    At leading order, the boosted temperature contrast is 
    \begin{align}
        \tilde\delta_{\rm T}({\bs{n}}) = \delta_{\rm T}(\bs{n}) +{\cal D}_{\rm CMB}\, \bs{n}\cdot {\bs{v}}_o^{\rm CMB}, 
    \end{align}
    where $\bs{n}$ is a unit vector in the direction of observation.  Using units with $c=1$, the dimensionless CMB dipole factor is ${\cal D}_{\rm CMB}=1$.
    
    The CP requires the large-scale distribution of sources on the sky to have the same dipole -- i.e., the relative velocity $\bs{v}_o$ extracted from these sources should agree with that from the CMB as
    \begin{align}\label{vgvc}
        \bs{\hat{v}}_{o} \approx \bs{\hat{v}}_o^{\rm CMB}, ~~{v}_{o} \approx {v}_o^{\rm CMB}.
    \end{align}
    The boosted two-dimensional (2D) number density contrast is
    \begin{align}\label{eq:2Dboosted}
        \tilde\delta_{2\rm D}({\bs{n}}) = \delta_{2\rm D}(\bs{n}) +{\cal D}\, \bs{n}\cdot {\bs{v}}_{o}.
    \end{align}    
    The dimensionless  number count dipole factor ${\cal D}$ is independent of $v_o$, but it does depend on the sample.

    \citet{ellis1984} estimated the dipole factor for the 2D projection of radio sources as
    \begin{equation}\label{eq:EB}
        {\cal D}_{\rm EB} = 2 + x\,(1+\alpha),
    \end{equation}
    in which the power-law indices $x$ and $\alpha$ are constants that determine the number counts per solid angle (${\cal N}_\Omega \propto S_c^{-x}$, with $S_c$ being the flux threshold of the survey) and the flux density of sources in their rest frame ($S(\nu_\mathrm{obs}) \propto \nu_\mathrm{obs}^{-\alpha}$) at fixed observed frequency $\nu_\mathrm{obs}$.\,This is potentially the most powerful consistency test of the CP, but it faces major observational obstacles. 
    
    First, the galaxy sample must cover a wide sky area and a deep redshift range, with a high number density. High redshifts are required to extract the cosmic dipole and the lowest-redshift sources need to be removed to avoid nonlinear contamination \citep{tiwari2016,bengaly2019}. Second, measurements of large-scale features face well-known systematic errors (effects of the mask, instrumental/survey systematics, stellar contamination, etc.). In this paper, we focus on a key theoretical systematic error that arises when estimating the dimensionless dipole factor ${\cal D}$ without using radial information.
    
    Different groups have applied  Equation \eqref{eq:EB} to extract the relative velocity from the measured dipole in wide-area radio continuum surveys \citep[see, e.g.,][]{blake2002,gibelyou2012,rubart2013,tiwari2015,ghosh2016,colin2017,bengaly2018}. A general conclusion of these analyses is that the dipole direction is typically consistent with that of the CMB, but the relative velocity amplitude is  significantly larger:\footnote{Note that \citet{darling2022} finds no tension of this form.}
    \begin{eqnarray}\label{cpeb}
    \bs{\hat{v}}_o^{\rm EB} \approx \bs{\hat{v}}_o^{\rm CMB},~~~ v_o^{\rm EB} > v_o^{\rm CMB}.
    \end{eqnarray}

    Recently, one of the most thorough analyses has used the very large CatWISE2020 sample of quasars to apply the Ellis--Baldwin test \citep{secrest2021,secrest2022}. Their conclusion reinforces Equation \eqref{cpeb}, finding a tension with the CMB velocity amplitude at $\sim 5\sigma$. 
    
    Some works have attempted to resolve this tension by invoking superhorizon effects \citep[for example, see][]{Das:2021ssc,Domenech:2022mvt,Tiwari:2021ikr}. In this work, we do not investigate alternative cosmological models.   Instead, we consider a possible theoretical systematic in the velocity estimates. We extend the pioneering analysis of \citet{dalang2022}, who showed that the presence of parameter evolution can lead to significant corrections on the Ellis--Baldwin approximation. We investigate the impact of different quasar luminosity function (QLF) models on the predicted amplitude of the kinematic dipole. In Section \ref{sec:review}, we review the theoretical predictions for the expected kinematic dipole amplitude in three dimensions, ${\cal D}_{3\rm D}$, and how it can be projected into the 2D dipole factor ${\cal D}$ of Equation \eqref{eq:2Dboosted}. In Section \ref{sec:data}, we introduce the QLF data and the theoretical modelling used for the fits. The Markov Chain Monte Carlo (MCMC) analysis and the goodness-of-fit criterion are described in Section \ref{sec:methods}. The results and conclusion are presented in Section \ref{sec:results} and Section \ref{sec:conclusions}, respectively. 

\section{Correction to the standard dipole}\label{sec:review}

In all of the results characterised by Equation \eqref{cpeb}, the dipole is measured and then Equation \eqref{eq:EB} is used to extract the relative speed. However, ${\cal D}_{\rm EB}$ assumes that radial information can be neglected. In reality, ${\cal D}$ is a radial projection of the three-dimensional redshift-dependent dipole factor \citep{maartens2018}
\begin{equation}\label{eq:3dDip}
 {\cal D}_{3\rm D}= 2+ \frac{\dot{\cal H}}{{\cal H}^2} +\frac{2}{{r}{\cal H}}-\frac{5s}{r {\cal H}}- b_e,
\end{equation}which can be derived from linear perturbations.\footnote{{There is also an intrinsic dipole sourced by primordial perturbations \citep{tiwari2016,nadolny2021}, which is not relevant for our discussion.}}

The first three terms on the right are purely cosmological ($r$ is the comoving line-of-sight distance) and the last two contain the magnification ($s$) and evolution ($b_e$) biases, which depend on the sample's luminosity function $\Phi$ through the background comoving number density (in the source frame)
\begin{eqnarray}\label{eq:nbar}
    n(z,M_{\rm c})=\int_{-\infty}^{M_{\rm c}(z)}{\rm d}M\,\Phi(z,M).
\end{eqnarray}
The absolute magnitude of a source (corresponding to its intrinsic luminosity) is given by
\begin{eqnarray}\label{eq:magm}
    M = m(z) - \mu(z) - K(z),
\end{eqnarray}
where $m$ is the apparent magnitude  measured at redshift $z$ (corresponding to observed flux), $\mu$ is the distance modulus (defined by the background luminosity distance) and $K$ is the $K$-correction. The apparent magnitude cut of the survey $m_{\rm c}$ leads to an absolute threshold $M_{\rm c}(z)$ that depends on redshift via Equation \eqref{eq:magm}.  The magnification and evolution biases become \citep[][]{challinor2011,alonso2015,maartens2021}
\begin{align} \label{eq:magbias}
    s(z,M_{\rm c}) &= \frac{\partial \log {n}(z,M_{\rm c})}{\partial  M_{\rm c}}= {\frac{1}{\ln 10}\, \frac{\Phi(z,M_{\rm c})}{{n}(z,M_{\rm c})},}
    \\ \label{eq:evolbias}
    b_e(z,M_{\rm c}) &= -\frac{\partial \ln {n}(z,M_{\rm c})}{\partial \ln(1+z)}
    \notag\\
    &= {-\frac{(1+z)}{{n}(z,M_{\rm c})}
    \int_{-\infty}^{M_{\rm c}(z)}{\rm d}M\, \frac{\partial \Phi(z,M)}{\partial z}\bigg|_{M},}
\end{align}
in which $s$ determines whether sources will be included or excluded from the sample due to lensing convergence and $b_e$ describes the deviation of the comoving number density of sources from the conserved case ($b_e=0$). Note the partial redshift derivative in $b_e$ must be taken at fixed magnitude $M_{\rm c}$.

The observed source number density ${\cal N}$ (number per redshift per solid angle), is not the same as the number density $n$ (number per comoving volume) measured at the source \citep{Bonvin:2011bg,challinor2011,alonso2015,maartens2018,maartens2021}. At the background level, these quantities are related by
\begin{eqnarray}\label{ncaln}
    {\cal N}=\frac{{\rm d}{N}}{{\rm d}z\,{\rm d}\Omega}
=\frac{r^2}{(1+z)\mathcal{H}}\,n,
\end{eqnarray}
where ${N}$ is the number of sources, which is the same in the observer and source frames, and ${\rm d}{N}={\cal N} {\rm d}z\,{\rm d}\Omega=n \,{\rm d}V$, with $V$ the comoving volume. The observed number density projected on the sky becomes
\begin{eqnarray}\label{nom}
    {\cal N}_\Omega =\int_0^\infty {\rm d}z\,{\cal N}(z).
\end{eqnarray}

The 2D dipole factor is found by projecting Equation \eqref{eq:3dDip} along the radial direction, weighted by the number counts \citep{nadolny2021,dalang2022}\footnote{In these papers, their ${\cal D}_{\rm kin}$ equals our ${\cal D}\,v_o$.}:
\begin{equation}
    \label{d2d3}
{{\cal D} = \int_0^\infty
{\rm d}z\, f(z)\,{\cal D}_{3\rm D}(z)~~ \mbox{where}~~ f(z)= \frac{{\cal N}(z)}{{\cal N}_\Omega}.}
\end{equation}
The result is
\begin{equation}
\label{eq:Dkin}
       {{\cal D} = {\cal D}_{\rm cosmo}+{\cal D}_{\rm mag}+{\cal D}_{\rm evol},}
\end{equation}
where
\begin{align}
    {\cal D}_{\rm cosmo} &= \int_0^\infty{\rm d}z\,f\Big[2+\frac{2}{{r}{\cal H}} + \frac{\dot{\cal H}}{{\cal H}^2}\Big],\label{eq:dcosmo}\\
    {\cal D}_{\rm mag} &= -2\int_0^\infty{\rm d}z\,f\,\frac{x}{{r}{\cal H}}~~\mbox{with} ~~x=\frac{5}{2}\,s,
    \label{eq:dmag}\\
    {\cal D}_{\rm evol} &= - \int_0^\infty{\rm d}z\,f\,b_e .\label{eq:devol}
\end{align}
For a constant magnification bias, $x$ corresponds to the Ellis--Baldwin parameter in Equation \eqref{eq:EB}, but in general it evolves with redshift. The dipole factor ${\cal D}$, in Equations \eqref{eq:Dkin} -- \eqref{eq:devol}, depends on the evolution bias $b_e$, which is unique for the sample, but not on the variable source spectral index $\alpha$ \citep[][]{carballo1999, secrest2021}. This important feature arises from the fact that the absolute magnitude cut, $M_{\rm c}(z)$, does not depend on $\alpha$ at a constant redshift, but at a fixed distance $r$ \citep[][]{dalang2022}.

The Ellis--Baldwin formula is an approximation to Equation \eqref{d2d3}, ${\cal D}_{\rm EB}={\cal D}+\Delta {\cal D}$, that includes a correction $\Delta {\cal D}$ which does not affect the dipole direction. The dipole amplitude is extracted directly from data and is equal to the dipole factor times the relative speed
\begin{eqnarray}\label{cpeb2}
{\cal D}_{\rm EB}\,v_{o}^{\rm EB} = {\cal D}\,v_o.
\end{eqnarray}When using the exact formula (Equation \ref{d2d3}) for the dipole factor ${\cal D}$, the correct relative speed $v_o$ is extracted from the measurements. However, using the Ellis--Baldwin approximation ${\cal D}_{\rm EB}$ we extract an estimate of the relative speed that may differ from $v_o$: $v_o^{\rm EB}=v_o+\Delta v_o$. For example, if the relative speed is overestimated, i.e., $v_{o}^{\rm EB}>v_o$, then ${\cal D}_{\rm EB}$ underestimates ${\cal D}$, since ${\Delta v_o}/{v_o}=-{\Delta {\cal D}}/{\cal D}$ by Equation \eqref{cpeb2}.

\section{Data}\label{sec:data}

    The QLF measurements used in this work are from \citet{palanque2016} (PD2016 hereafter) and include 13876 quasars: 7900 were identified through flux variability \citep[][]{schmidt2010, palanque2011} in the Sloan Digital Sky Survey (SDSS) photometric $\lbrace u, g, r, i, z \rbrace$ bands and 5976 were spectroscopically identified in the Stripe 82 region. The data is part of the extended Baryon Oscillation Spectroscopic Survey (eBOSS) in SDSS-IV and is available in Table~A.1 of \citetalias{palanque2016}. The magnitudes have been corrected for Galactic extinction with a magnitude limit of $g_{\rm dered} = 22.5$ in the dereddened $g$ band.    
    
    \subsection{Theoretical modelling}\label{ssec:models}
    
    The number counts of quasars have two main sources of uncertainty: the bright end has a small number of objects, while the faint end is limited by the survey strategy. The large uncertainties on both ends of the luminosity function leads us to the adoption of a double power-law fit for the comoving space density of quasars \citep{boyle2000,croom2004,croom2009,ross2013}
    \begin{equation}\label{eq:PhiDB}
        \Phi(z,M_g) = \frac{\Phi_*}{10^{0.4(a+1)(M_g-M_*)}+10^{0.4(b+1)(M_g-M_*)}}.
    \end{equation}Above, $\Phi_*$ is a normalisation factor related to the characteristic number density of quasars, $M_*$ is the break magnitude, associated with their characteristic luminosity, $a$ and $b$ describe, respectively, the behaviour of the bright and faint ends, and
    \begin{equation}\label{eq:MgKz}
        M_g(z) = m_g - \mu(z) - [K(z)-K(z=2)]
    \end{equation}is the absolute magnitude corresponding to the observed apparent magnitude $m_g$ in the $g$ band after the $K$-correction
    \begin{equation}
        K(z) = -2.5(1+\alpha_v)\log_{10}(1+z), \,\,\, \alpha_v \simeq -0.5
    \end{equation}has been applied \citepalias[][]{croom2009,palanque2016}.
    
    The number density of quasars has a strong redshift dependence and the QLF may evolve in different ways: in the pure-luminosity evolution (PLE) case, the characteristic number density remains constant, but the break magnitude evolves over time; in the pure-density evolution (PDE), the luminosity of individual sources remain constant, while the number density varies with time; or it could be combination of both luminosity and density evolution (LEDE). Redshift dependence may also impact the bright- and faint-end slopes.
    
    Since the QLF is very uncertain, this work aims at investigating different evolution models for $\Phi_*, M_*, a$ and $b$, and to assess their impact on the kinematic dipole ${\cal D}$. The models are described in detail in Appendix\,\ref{ap:models}. They are \citepalias{croom2004,ross2013,palanque2016}:
    \begin{itemize}
        \item PLE -- the QLF redshift evolution comes from the break magnitude $M_*$, with the pivot scale $z_p=2.2$ allowing the bright and faint slopes to vary for low ($z<z_p$) and high ($z>z_p$) redshifts.
        \item LEDE -- both luminosity and number density vary with time.\,We consider a linear (LEDE$_7$) and quadratic (LEDE$_8$) redshift dependence on the break magnitude $M_*(z)$ and their extension, in which both slopes evolve linearly with redshift (LEDE$_{7+2}$ and LEDE$_{8+2}$).
        \item PLE+LEDE -- it combines the PLE functional form for $z<z_p$, and the LEDE$_7$ model for $z>z_p$. We allow the bright slope $a$ to evolve with redshift. 
    \end{itemize}

\section{Method}\label{sec:methods}

    \subsection{Maximum likelihood}\label{ssec:MCMC}
    
        We consider a maximum-likelihood approach to sample the parameter space of the models of Section \ref{ssec:models}. Because the QLF is estimated from the number counts of quasars inside each magnitude bin, its error can be approximately modelled by Poissonian error bars. Therefore, we model the log-likelihood function as \citep{pozzetti2016}
        \begin{equation}\label{eq:likelihood}
            \ln {\cal L} = \sum_{i,j}\frac{\Delta^2_{i,j}}{\sigma_{i,j}^2},
        \end{equation}In this equation, $\Delta^2_{i,j} \equiv 1 - \Phi_\theta(z_i,M_j)/\Phi_{\rm obs}(z_i,M_j) + \ln[\Phi_\theta(z_i,M_j)/\Phi_{\rm obs}(z_i,M_j)]$, $\sigma_{i,j}^2 = 1/N_{\rm obs}(z_i,M_j)$, $\Phi_\theta$ is given by the model of Equation (\ref{eq:PhiDB}). Finally, $\Phi_{\rm obs}$ and $N_{\rm obs}$ are, respectively, the corresponding QLF measurements and the angle averaged number counts in each redshift $z_i$ and magnitude $M_j$ bin. To maximise Equation (\ref{eq:likelihood}), we used the MCMC sampler \texttt{emcee} \citep{emcee} with flat priors for the fitting parameters.
    
    \subsection{Goodness of fit}
    
        To assess the goodness of fit of the models considered in Section \ref{ssec:models} to the eBOSS data, in Table~\ref{tab:MCMCresults} we present the Bayesian information criterion (BIC)
        \begin{equation}
            {\rm BIC} = p\ln(n) - 2\ln{\cal L}^*,\label{eq:BIC}
        \end{equation}in which $p$ is the number of parameters in the model, $n$ is the number of data points used, and $\ln{\cal L}^*$ is the log-likelihood computed at the best-fitting values \citep[e.g.][]{liddle2004}.

\section{Results}\label{sec:results}

    \begin{figure*}[!t] 
        \centering
        \vspace{1em}
        \includegraphics[width=\textwidth]{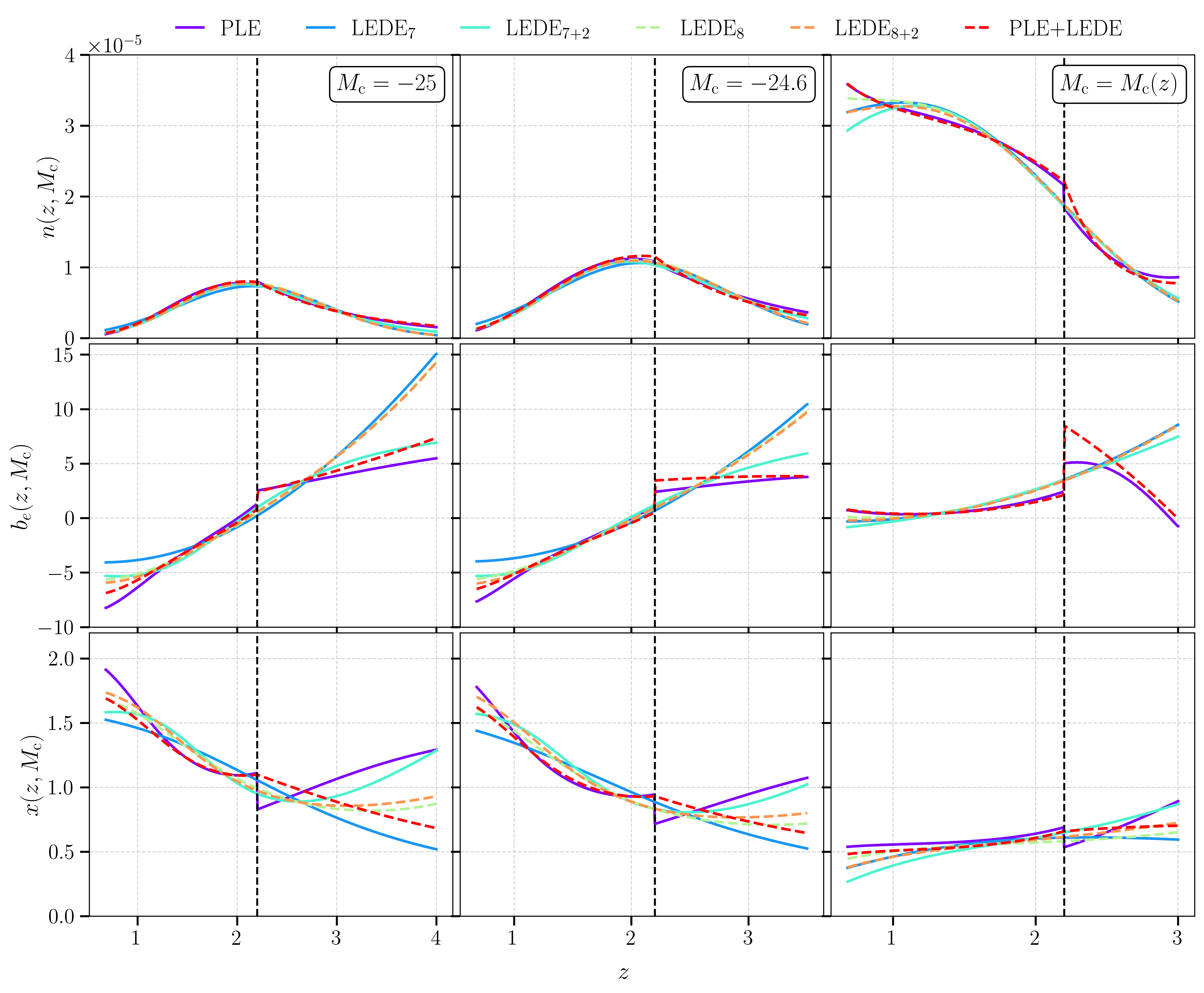}
        \caption{Comoving number density (\textit{top}), evolution (\textit{middle}) and magnification (\textit{bottom}) biases derived from the QLF with different absolute magnitude thresholds, assuming Planck 2015 cosmology \citep[][]{planck2015}. The pivot redshift, $z_p=2.2$, entering the PLE and PLE+LEDE models is indicated by the vertical dashed lines.}
        \label{fig:derived_quantities}
        \vspace{2em}
    \end{figure*}
    
    \begin{table*}[!t]
    \centering
    \caption{1$\sigma$ constraints for ${\cal D}_{\rm kin} = {\cal D}\, v_o^{\rm CMB}/c$, in $10^{-3}$ units, obtained from a random subset of 5000 MCMC samples, with Planck 2015 and $w$CDM cosmologies. We present the results with different absolute magnitude thresholds of $-25$, $-24.6$ and a varying threshold $M_{\rm c}(z)$. The effect of neglecting the redshift evolution of the magnification and evolution biases (Equation \ref{eq:effective_values}) is shown in the last column for the Planck cosmology and absolute magnitude threshold $M_{{\rm c},g} = -25$.}
    \label{tab:results}
    \hspace{-2cm}\begin{tabular}{cc|ccc|cc}
         & \multicolumn{6}{c}{${\cal D}_{\rm kin} \,\, [10^{-3}]$} \\ \cline{2-7}
         & $w$CDM & \multicolumn{3}{c}{Planck 2015} & \multicolumn{2}{c}{$+\,\lbrace b_e^{\rm eff}, s^{\rm eff}\rbrace$} \\ \cline{2-7}
      & \multirow{2}{*}{$M_{{\rm c},g} = -25$}   & \multirow{2}{*}{$M_{{\rm c},g} = -25$}    & \multirow{2}{*}{$M_{{\rm c},g} = -24.6$} &  \multirow{2}{*}{$M_{{\rm c},g} = M_{{\rm c},g}(z)$}   &   \multicolumn{2}{c}{$M_{{\rm c},g} = -25$} \\
      & & & & & Best fit & Chains \\
      \cline{1-7}
        PLE          & $0.72^{+0.10}_{-0.10}$ & $0.75^{+0.10}_{-0.10}$ & $1.42^{+0.10}_{-0.10}$ & $0.39_{-0.05}^{+0.05}$ & $0.81$ & $0.84_{-0.12}^{+0.12}$ \\
        LEDE$_7$     & $0.32^{+0.03}_{-0.03}$ & $0.34^{+0.03}_{-0.03}$ & $0.90^{+0.04}_{-0.04}$ & $0.53_{-0.06}^{+0.06}$ & $0.49$ & $0.28_{-0.04}^{+0.04}$ \\
        LEDE$_{7+2}$ & $0.59^{+0.06}_{-0.06}$ & $0.62^{+0.06}_{-0.06}$ & $1.14^{+0.06}_{-0.06}$ & $0.67_{-0.05}^{+0.05}$ & $0.72$ & $0.75_{-0.07}^{+0.07}$ \\
        LEDE$_8$     & $0.39^{+0.03}_{-0.03}$ & $0.42^{+0.03}_{-0.03}$ & $1.01^{+0.04}_{-0.04}$ & $0.44_{-0.06}^{+0.06}$ & $0.54$ & $0.38_{-0.04}^{+0.04}$ \\
        LEDE$_{8+2}$ & $0.40^{+0.03}_{-0.05}$ & $0.42^{+0.03}_{-0.05}$ & $0.99^{+0.04}_{-0.04}$ & $0.50_{-0.07}^{+0.07}$ & $0.53$ & $0.42_{-0.07}^{+0.05}$ \\
        PLE+LEDE     & $0.60^{+0.08}_{-0.08}$ & $0.62^{+0.08}_{-0.08}$ & $1.31^{+0.08}_{-0.09}$ & $0.47_{-0.05}^{+0.05}$ & $0.72$ & $0.77_{-0.11}^{+0.11}$ \\ \cline{1-7}
    \end{tabular}
    \vspace{2em}
    \end{table*}
    
    \begin{figure*}[!t]
        \centering
        \includegraphics[width=\textwidth]{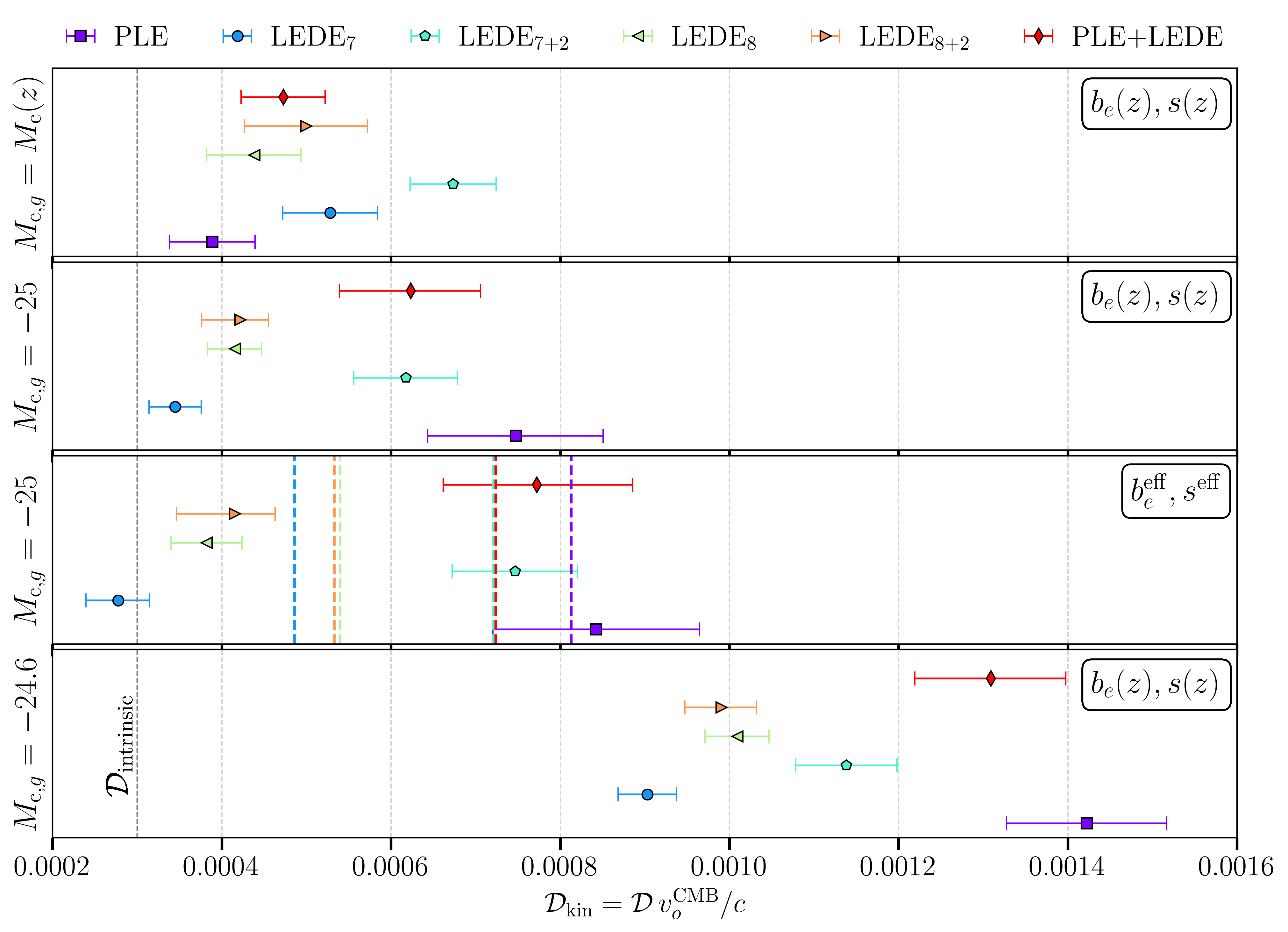}
        \caption{1$\sigma$ constraints for ${\cal D}_{\rm kin} = {\cal D}, v_o^{\rm CMB}/c$ obtained from the same random subset of 5000 MCMC samples used to derive the constraints of Table \ref{tab:results} with the $\Lambda$CDM cosmology. The grey dashed line is a reference for the intrinsic dipole \citep{dalang2022}. The top panel shows the analysis for a varying magnitude threshold $M_c(z)$, i.e., the corresponding absolute magnitude for $g = 22.5$ at each redshift used to compute $n(z)$. The second and third panels have a fixed threshold given by the last redshift bin (see Figure \ref{fig:all_models} for illustration), and the last panel neglects the last bin (i.e., $M_c(z)$ corresponds to $g = 22.5$ at $z=3.25$). The third panel shows the constraints after neglecting the redshift evolution (Equation \ref{eq:effective_values}) with $s(z)$ and $b_e(z)$ projected in redshift for each MCMC sample (\textit{points with error bars}), and for projections using the best-fitting functions (\textit{dashed lines}). Neglecting redshift evolution (\textit{third panel}) causes significant tensions in some cases, and neglecting data from the last redshift bin shifts the amplitude mainly due to the different redshift range used to obtain ${\cal D}$ (\textit{last panel}). Different QLF models can lead to a $\sim 3\sigma$ tension in ${\cal D}$.}
        \label{fig:dipole}
        \vspace{2em}
    \end{figure*}
    
    We fitted the QLF models described in Appendix \ref{ap:models} to the eBOSS data. In Table \ref{tab:results}, we show the mean of random subsets of 5000 samples, obtained from the complete sets of $\gtrsim$ 300000 MCMC chains used to derive the constraints on the QLF models, and the 1$\sigma$ constraints obtained for the amplitude of the kinematic dipole from those subsets.\footnote{We checked the constraints against different subsamples, and also by varying the number of MCMC chains in the subsets, finding no significant changes.} The posteriors for the QLF are presented in Table \ref{tab:MCMCresults}, together with their goodness of fit. 
    
    In Figure \ref{fig:derived_quantities}, the number density, evolution and magnification biases derived from the best-fitting QLF are shown, after assuming a flat $\Lambda$CDM model with parameters $h = 0.679, n_s = 0.9681, \sigma_8 = 0.8154, \Omega_m = 0.3065$ and $\Omega_b h^2 = 0.02227$ \citep{planck2015} (same cosmology employed in \citetalias{palanque2016} for the distance modulus $\mu$). To assess the dependence of our results on the fiducial cosmology used to compute Equation (\ref{eq:Dkin}), in Table \ref{tab:results} the constraints obtained with a flat $w$CDM cosmology with parameters $h=0.72, n_s = 0.963, \sigma_8 = 0.852, \Omega_m = 0.275, \Omega_b h^2 = 0.02258, \Omega_{\rm DE} = 0.725$, and $w = -1.2$ \citep{rasera2022} are also shown. The constraints are robust against this change as seen in Table \ref{tab:results}.
    
    We considered different absolute magnitude thresholds: $M_{{\rm c},g}(z_8) = -25$, corresponding to the apparent magnitude cut $g_{\rm dered}=22.5$ at the centre $z_8 = 3.75$ of the last redshift bin \citep[e.g.][]{wang2020}, $M_{{\rm c},g}(z_7) = -24.6$, corresponding to $g_{\rm dered}$ at $z_7 = 3.25$, and a varying $M_{{\rm c},g}(z)$ obtained from Equation (\ref{eq:MgKz}) at each redshift used to compute $n(z,M_{\rm c}), b_e(z,M_{\rm c})$, and $s(z,M_{\rm c})$.
    
    In Figure \ref{fig:dipole}, we show the impact of the magnitude cut $M_{\rm c}$ on the results (first, second and fourth panels). By fixing a threshold such as $M_{\rm c} = -25$, we discard data at lower redshifts (see Figure \ref{fig:all_models}). Hence, the number density increases by allowing larger values for $M_{\rm c}$. By considering $M_{\rm c} = -24.6$, we neglect the last redshift bin in the analysis to avoid adding sources too faint to be seen at the highest $z$ bin. Thus, we are probing the kinematic dipole with lower-$z$ sources and this shifts ${\cal D}$ to larger values as the integration range changes (fourth panel of Figure \ref{fig:dipole}). Most of the impact comes from the ${\cal D}_{\rm cosmo}$ term.
    
    The evolution bias derived from the LEDE$_7$, LEDE$_8$ and LEDE$_{8+2}$ models are very similar at high redshifts. While the evolution bias is rather consistent among the different models, the magnification bias has a larger dependence on the QLF. However, the contributions coming from Equation (\ref{eq:dmag}) exhibit a smaller dispersion due to the $1/r{\cal H}$ suppression in ${\cal D}_{\rm mag}$. The contributions from the evolution term (Equation \ref{eq:devol}) are more sensitive to the differences in the QLF. This is manifested in the final 2D amplitude ${\cal D}$: LEDE$_7$, LEDE$_8$ and LEDE$_{8+2}$ agree within $\sim 1\sigma$, while the LEDE$_{7+2}$, PLE and PLE+LEDE have $\sim 1\sigma$ consistency among each other. 
    
    We also tested the robustness of our results against the minimum redshift considered; the values presented in Table \ref{tab:results} were obtained by numerically integrating Equations (\ref{eq:dcosmo}) -- (\ref{eq:devol}) from $z_{\rm min}\approx 0$ to $z_{\rm max} = 4$ (the maximum redshift of the sample). Because the number of sources below $z = 0.68$ (minimum redshift) is very small, we find no significant changes in the constraints for the dipole amplitude. Still, it is worth stressing the fact that the intrinsic dipole is considerable for low-$z$ sources; therefore, in principle it is better to remove $z\lesssim 0.5$ to avoid nonlinear contamination in real data \citep[][]{tiwari2016,bengaly2019}.
    
    Finally, considering the Planck 2015 cosmology and $M_{{\rm c},g} = -25$, we neglect the redshift evolution of the magnification and evolution biases, i.e.,
    \begin{equation}\label{eq:effective_values}
        s^{\rm eff} = \int {\rm d}z\, f(z)s(z) \,\,\,\text{ and }\,\,\, b_e^{\rm eff} = \int {\rm d}z\, f(z) b_e(z).
    \end{equation}This has been done for (a) each one of the 5000 random chains, i.e., calculating $f(z)$, $s(z)$ and $b_e(z)$ for each chain and then projecting along $z$, and (b) by fixing them to the best-fitting functions. In the first case, the results agree within the 1$\sigma$ errors; for the second case, however, the amplitudes are overestimated by more than 3$\sigma$ for the LEDE$_7$ and LEDE$_8$ models.
    
    Our results summarise the expected dipole in the number counts of quasars from a pure kinematic origin. They are all above the intrinsic dipole expected from the clustering anisotropy (grey dashed line in Figure \ref{fig:dipole}), which comes from fluctuations in the number density around the FLRW metric \citep[][]{nadolny2021} and can be distinguished given the 1$\sigma$ errors.

\section{Conclusions} \label{sec:conclusions}

    In this work, we extended the analysis of \citet{dalang2022} to include uncertainties on the magnification $s$ and evolution $b_e$ biases (Equations \ref{eq:magbias} and \ref{eq:evolbias}). We used a maximum-likelihood approach on six QLF models (Equation \ref{eq:PhiDB}) to derive the number density $n(z)$ (Equation \ref{eq:nbar}), $s(z)$, and $b_e(z)$ \citep{wang2020}, finding that models of QLF have a non-negligible impact on the biases (Figure \ref{fig:derived_quantities}). From MCMC chains, we calculated the kinematic dipole amplitude ${\cal D}$ (Equation \ref{eq:Dkin}), finding that the QLF introduces a large model dependence on ${\cal D}$. In some cases, a $\sim 3\sigma$ tension on the dipole amplitude computed with different models (Figure \ref{fig:dipole}) is found.
    
    The constraints on ${\cal D}$ are robust to a change of cosmology (see Table \ref{tab:results}). Neglecting data from the last redshift bin does not degrade the constraints, as the $1\sigma$ error bars on ${\cal D}$ are of the same order, although the amplitude increases with higher magnitude thresholds $M_{{\rm c},g}$, mainly due to a different redshift range.
    
    We analysed the impact of neglecting the redshift evolution of the magnification and evolution biases by computing their effective values, $s^{\rm eff}$ and $b_e^{\rm eff}$ (Equation \ref{eq:effective_values}). When these parameters were computed for each MCMC sample, we found a shift in the expected amplitude within the $1\sigma$ region of the cases in which the redshift evolution was accounted for; by obtaining them from the best-fitting functions, we found an overestimation of the 2D amplitude ${\cal D}$ when redshift evolution was neglected in quantities that evolve with $z$. 
    
    The QLF introduces a model dependence in the dipole amplitude, contrasting significantly with the model-independent null hypothesis of \citet{ellis1984} for testing the validity of the CP. As we have shown, with no single description of quasar evolution and multiple models fitting the data equally well, testing the CP with the quasar dipole is complex and not as straightforward as assumed. The difficulty is exacerbated for radio continuum catalogues, which have no redshift information.

    Finally,  conclusions from our work cannot be directly extrapolated to the results of \citet{secrest2022}, given the different wavelengths, magnitude thresholds and redshift information for the CatWISE2020 and eBOSS catalogues. In addition, following \citet{dalang2022}, we used an alternative theoretical approach to \citet{secrest2022} (and to previous work on radio continuum samples) -- which is based on the evolution bias of the sample, rather than on the spectral indices of individual sources. We find, consistent with \citet{dalang2022}, that the magnification and evolution biases  vary with redshift for all the models considered for the eBOSS quasars, with significant implications for the dipole amplitude. Therefore it is paramount to investigate whether evolution effects are also present in the data considered for the analyses leading to Equation \eqref{cpeb}, before strong claims about violation of the CP are validated.

\vspace{-1em}
\section*{Acknowledgements}
We thank Charles Dalang, Carolina Queiroz, José Luis Bernal, Mike Shengbo Wang, and Phil Bull for discussions and comments. C.G. and C.C. are supported by the UK Science \& Technology Facilities Council consolidated grant ST/T000341/1. J.P. received support from the French government under the France 2030 investment plan, as part of the Initiative d'Excellence d'Aix-Marseille Université -A*MIDEX (AMX-19-IET-008). R.M. is supported by the South African Radio Astronomy Observatory and the National Research Foundation (grant No. 75415).

\facilities{This work uses data from the extended Baryon Oscillation Spectroscopic Survey of the Sloan Digital Sky Survey (SDSS-IV/eBOSS), as extracted from Table~A.1 of \citetalias{palanque2016}.}

\software{We made extensive use of \href{https://emcee.readthedocs.io/}{\texttt{emcee}} \citep{emcee}, \href{https://github.com/lesgourg/class_public}{\texttt{CLASS}} \citep{blas2011}, \texttt{numpy} \citep{harris2020}, \texttt{scipy} \citep{2020SciPy}, and \texttt{matplotlib} \citep{hunter2007}.}



\appendix

\section{Quasar luminosity function models}\label{ap:models}

In this appendix we describe the QLF models considered for the main analysis.

\subsection{Pure-Luminosity Evolution (PLE)}\label{sssec:PLE}
    In this model, the luminosity function evolves solely through a redshift evolution in the break magnitude $M_*$ \citepalias{boyle2000, palanque2016}
    \begin{equation}
        M_*(z) = M_*(z_p) - 2.5\left[k_1(z-z_p) + k_2(z-z_p)^2\right].
    \end{equation}
    
    Early studies showed a preference for this model, but deeper data revealed deviations from the PLE model at redshifts larger than $z\gtrsim 2$ \citep[][]{ross2013} for $z_p = 0$. Therefore, the bright- and faint-end slopes are allowed to vary for low ($z<z_p$) and high ($z>z_p$) redshifts, $a_{\lbrace l,h\rbrace},b_{\lbrace l,h\rbrace}$, and so are $k_1$ and $k_2$. The pivot redshift is $z_p = 2.2$. In this model, there are 10 free parameters: $\lbrace a_l,b_l,a_h,b_h,k_{1,l},k_{2,l},k_{1,h},k_{2,h},M_*(z_p),\log_{10}\Phi_* \rbrace$ \citepalias{palanque2016}.
    
    \subsection{Luminosity and Density Evolution (LEDE)}
    
    We also consider the case in which both the luminosity and number density vary with time: the so-called Luminosity Evolution + Density Evolution (LEDE) model \citepalias{ross2013, palanque2016}. We analyse this particular case motivated by the fact that the discontinuity present at the pivot redshift $z_p = 2.2$ for the derived quantities is removed, thus leading to a smooth behaviour with redshift for the derived quantities. 
    
    We will take the density and magnitude evolution, respectively, as
    \begin{equation}
        \log_{10}\Phi_*(z) = \log_{10}\Phi_*(z_p) + c_{1}(z-z_p) + c_{2}(z-z_p)^2,\label{eq:PDE1}
    \end{equation}
    \begin{equation}
        M_*(z) = M_*(z_p) + c_3(z-z_p).
    \end{equation}
    
    Because it has seven free parameters $\lbrace a,b$, $c_{1}, c_{2}, c_3$, $\log_{10}\Phi_*(z_p)$, $M_*(z_p)\rbrace$, we will label it LEDE$_7$ hereafter. Notice that, in this case, the bright and faint ends are fixed for the whole redshift range: $a(z) = a$, $b(z) = b$. 
    
    We shall also consider a quadratic redshift dependence for the break magnitude \citep{boyle2000,croom2009}
    \begin{equation}
        M_*(z) = M_*(z_p) + c_{3}(z-z_p) + c_{4}(z-z_p)^2.
    \end{equation}We dub this case LEDE$_8$ as it contains eight free parameters: $\lbrace a,b$, $c_{1}, c_{2}, c_{3}, c_{4}$, $\log_{10}\Phi_*(z_p)$, $M_*(z_p)\rbrace$. Again, $a$ and $b$ are fixed.
    
    Finally, we consider extensions to LEDE$_7$ and LEDE$_8$, in which both slopes are allowed to evolve linearly with redshift
    \begin{equation}
        a(z) = a(z_p) + c_a(z-z_p), \,\,\,\, b(z) = b(z_p) + c_b(z-z_p).
    \end{equation}These will be called, respectively, LEDE$_{7+2}$ and LEDE$_{8+2}$, where the $+2$ indicates the extra $c_a$ and $c_b$ parameters introduced for the redshift evolution in the bright and faint ends.
    
    \subsection{Composite Evolution Model (PLE+LEDE)}
    
    We also consider the combination of the PLE functional form for $z<z_p$, and the LEDE$_7$ model for $z>z_p$ \citepalias{palanque2016}. Other combinations with LEDE$_8$ and redshift dependencies of the slopes are possible. However, we focus on this particular combination of PLE + LEDE$_7$ and allow only the bright slope to evolve with redshift: $\alpha(z) = \alpha(z_p) + c_a(z-z_p)$. In total, we have 10 free parameters: $\lbrace \alpha(z_p), \beta, k_1, k_2, c_{1}, c_{2}, c_3, c_a, \log_{10}\Phi_*(z_p), M_*(z_p)\rbrace$. 

\section{MCMC results}\label{ap:MCMC}

In Table~\ref{tab:MCMCresults}, we present the best-fitting values for the QLF obtained for the six models described in Section \ref{ssec:models}, from the MCMC analysis discussed in Section \ref{ssec:MCMC} by considering all eight redshift bins, removing the last bin and neglecting data from the last two bins.

There is a small impact in removing high-redshift information from the MCMC analysis for the LEDE$_7$ model: the parameters $c_1$, $c_2$ and $c_3$ show a slight degradation in the constraints, while the best-fitting values are in agreement within the $2\sigma$ region. The same happens for the PLE analysis, except for the high-redshift parameters, which are severely degraded with the removal of the last two redshift bins (i.e., considering data between $0.68 < z < 3.5$), as it is expected since there is no constraining power at the high-$z$ end, even though the fits are improved due to the fact that there are more data at low $z$ (which can also be seen by the decrease in the BIC value). We can also observe larger errors for the PLE+LEDE $c_1, c_2, c_3$ and $c_a$ parameters, with a $2\sigma$ disagreement for $c_1$ and $c_2$ after removing the last two redshift bins (considering only the data between $0.68 < z < 3$). On the other hand, LEDE$_8$, LEDE$_{8+2}$ and LEDE$_{7+2}$ are reasonably insensitive to high-redshift information, showcasing similar constraints for the three redshift ranges considered. 

\begin{figure*}[!t]
    \centering
    \includegraphics[width=\textwidth]{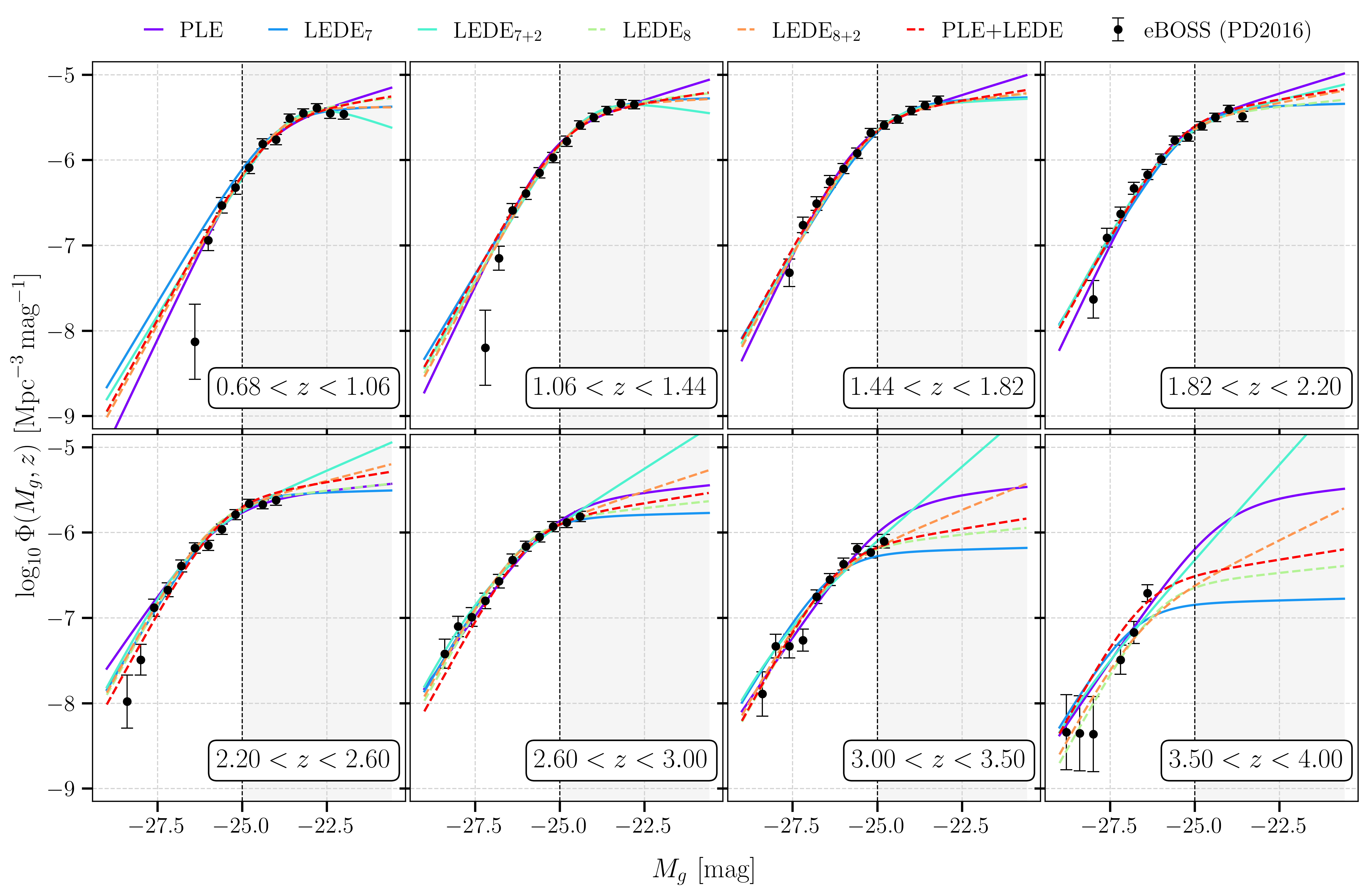}
    \caption{Best-fitting QLFs for the models described in Section \ref{ssec:models}. The corresponding best-fitting parameters are shown in Table~\ref{tab:MCMCresults}.\,The data points and uncertainties are taken from \citetalias{palanque2016}.\,The dashed line separates the region below the limiting magnitude cut $M_\mathrm{c} = -25$ for the last redshift bin \citep[][]{wang2020}.\,The shaded region is neglected when we consider all eight redshift bins for the dipole analysis.}
    \label{fig:all_models}
    \vspace{2em}
\end{figure*}

In Figure \ref{fig:all_models}, we show the best-fitting QLF from the corresponding MCMC analysis with all eight redshift bins. Dashed lines mark the limiting magnitude cut $M_\mathrm{c} = -25$ for the last redshift bin. When we consider all redshift bins and a constant magnitude cut for the dipole analysis, we neglect data contained in the shaded region. Discarding higher bins in the dipole analysis is equivalent to moving the dashed line and region of exclusion to fainter magnitudes.

\setlength{\LTcapwidth}{\textwidth}

\begin{longtable}[!t]{c|ccccccccc}
\caption{Best-fitting values for the parameters of the models described in Section \ref{ssec:models}, obtained from the MCMC analysis described in Section \ref{ssec:MCMC}. The last column gives the Bayesian Information Criterion (BIC, Equation \ref{eq:BIC}) for each analysis.}\label{tab:MCMCresults}\\
\hline 
\textbf{Model} & $\boldsymbol{z}$\textbf{-range} & \multicolumn{7}{c}{\textbf{Parameters}}  & \multicolumn{1}{c}{\textbf{BIC}} \\ \hline 
\endfirsthead
\multicolumn{10}{c}%
{\tablename\ \thetable\ -- \textit{Continued from previous page}} \\
\endhead
\multicolumn{10}{r}{\textit{Continued on next page}} \\
\endfoot
\hline
\endlastfoot
\multirow{18}{*}{\textbf{PLE}}  & \textbf{8 bins}  & $M_*(z_p)$ & $\log_{10}\Phi(z_p)$ & & & & & &  \multirow{6}{*}{176.97} \\
        & $0.68-4.0$ & $-25.86^{+0.12}_{-0.12}$ & $-5.63^{+0.04}_{-0.05}$ &  &  &  &  \\
        &            & $a_l$  & $b_l$   & $k_{1,l}$   & $k_{2,l}$  & &   \\
        & $0.68-2.2$ & $-3.09^{+0.09}_{-0.10}$  & $-1.31^{+0.04}_{-0.04}$ & $-0.16^{+0.05}_{-0.05}$ & $-0.42^{+0.04}_{-0.04}$ & & \\
        &            & $a_h$  & $b_h$ & $k_{1,h}$  & $k_{2,h}$  &  & \\
        & $2.2-4.0$  & $-2.46^{+0.05}_{-0.05}$  & $-1.10^{+0.07}_{-0.06}$ & $-0.42^{+0.07}_{-0.07}$ & $0.01^{+0.05}_{-0.05}$  & & \\ 
        \cline{2-10}
        & \textbf{7 bins} & \textbf{$M_*(z_p)$} & $\log_{10}\Phi(z_p)$ & & & & & & \multirow{6}{*}{ 171.88 } \\
        & $0.68-3.5$ & $-25.82^{+0.13}_{-0.12}$ & $-5.62^{+0.05}_{-0.05}$ & & & & \\
        &            & $a_l$ & $b_l$ & $k_{1,l}$ & $k_{2,l}$ & & \\
        & $0.68-2.2$ & $-3.06^{+0.09}_{-0.10}$  & $-1.29^{+0.04}_{-0.04}$ & $-0.16^{+0.05}_{-0.05}$ & $-0.42^{+0.04}_{-0.04}$ & & \\
        &            & $a_h$ & $b_h$ & $k_{1,h}$ & $k_{2,h}$ & & \\
        & $2.2-3.5$  & $-2.43^{+0.05}_{-0.06}$  & $-1.09^{+0.07}_{-0.07}$ & $-0.47^{+0.10}_{-0.10}$ & $0.05^{+0.08}_{-0.08}$  & & \\ 
        \cline{2-10}
        & \textbf{6 bins} & \textbf{$M_*(z_p)$} & $\log_{10}\Phi(z_p)$    & & & & & & \multirow{6}{*}{ 151.77 } \\
        & $0.68-3.0$ & $-25.63^{+0.14}_{-0.14}$ & $-5.53^{+0.05}_{-0.06}$ & & & & \\
        &            & $a_l$ & $b_l$ & $k_{1,l}$ & $k_{2,l}$ & & \\
        & $0.68-2.2$ & $-2.93^{+0.09}_{-0.10}$  & $-1.21^{+0.05}_{-0.05}$ & $-0.09^{+0.06}_{-0.06}$ & $-0.37^{+0.04}_{-0.04}$ & & \\
        &            & $a_h$ & $b_h$ & $k_{1,h}$ & $k_{2,h}$ & & \\
        & $2.2-3.0$ & $-2.37^{+0.07}_{-0.07}$  & $-0.96^{+0.12}_{-0.10}$ & $-1.03^{+0.24}_{-0.26}$ & $0.90^{+0.34}_{-0.33}$  & & \\ 
\hline
\multirow{18}{*}{\textbf{PLE+LEDE}} & \textbf{8 bins} & $M_*(z_p)$ & $\log_{10}\Phi(z_p)$ & $\alpha(z_p)$           & $\beta$ & & & & \multirow{6}{*}{175.50} \\
      & $0.68-4.0$ & $-25.52^{+0.16}_{-0.16}$ & $-5.48^{+0.06}_{-0.06}$ & $-2.80^{+0.08}_{-0.09}$ & $-1.16^{+0.06}_{-0.06}$ & & \\
      &            & $k_1$ & $k_2$ & & & & \\
      & $0.68-2.2$ & $-0.10^{+0.05}_{-0.05}$  & $-0.38^{+0.04}_{-0.04}$ & & & & \\
      &            & $c_1$ & $c_2$ & $c_3$ & $c_a$ & & \\
      & $2.2-4.0$ & $-0.62^{+0.06}_{-0.06}$  & $-0.06^{+0.05}_{-0.05}$ & $-0.67^{+0.13}_{-0.13}$ & $-0.01^{+0.11}_{-0.11}$ & & \\ 
      \cline{2-10}
  & \textbf{7 bins} & $M_*(z_p)$  & $\log_{10}\Phi(z_p)$ & $\alpha(z_p)$ & $\beta$ & & & & \multirow{6}{*}{ 163.30 } \\
  & $0.68-3.5$ & $-25.53^{+0.16}_{-0.16}$ & $-5.47^{+0.06}_{-0.06}$ & $-2.80^{+0.09}_{-0.09}$ & $-1.15^{+0.07}_{-0.06}$ & & \\
  &            & $k_1$ & $k_2$ & & & & \\
  & $0.68-2.2$ & $-0.08^{+0.05}_{-0.05}$  & $-0.37^{+0.04}_{-0.04}$ & & & & \\
  &            & $c_1$ & $c_2$ & $c_3$ & $c_a$ & & \\
  & $2.2-3.5$ & $-0.73^{+0.07}_{-0.07}$  & $0.07^{+0.07}_{-0.07}$  & $-0.69^{+0.16}_{-0.15}$ & $0.09^{+0.13}_{-0.13}$  & & \\ 
  \cline{2-10}
  & \textbf{6 bins} & $M_*(z_p)$ & $\log_{10}\Phi(z_p)$    & $\alpha(z_p)$ & $\beta$ & & & & \multirow{6}{*}{ 140.38 } \\
  & $0.68-3.0$ & $-25.58^{+0.16}_{-0.16}$ & $-5.47^{+0.06}_{-0.06}$ & $-2.84^{+0.09}_{-0.10}$ & $-1.16^{+0.07}_{-0.06}$ & & \\
  &            & $k_1$ & $k_2$ & & & & \\
  & $0.68-2.2$ & $-0.03^{+0.06}_{-0.06}$  & $-0.34^{+0.04}_{-0.04}$ & & & & \\
  &            & $c_1$ & $c_2$ & $c_3$ & $c_a$ & & \\
  & $2.2-3.0$ & $-1.08^{+0.12}_{-0.12}$  & $0.68^{+0.18}_{-0.18}$  & $-0.77^{+0.28}_{-0.26}$ & $0.36^{+0.22}_{-0.22}$  & & \\ \hline        
\multirow{12}{*}{\textbf{LEDE}$_\mathbf{7}$} & \textbf{8 bins} & $M_*(z_p)$ & $\log_{10}\Phi(z_p)$ & & & & & & \multirow{4}{*}{169.05}\\
 & \multirow{3}{*}{$0.68-4.0$} & $-25.37^{+0.16}_{-0.16}$ & $-5.46^{+0.05}_{-0.05}$ & & & & & &                               \\
 &                  & $a$ & $b$ & $c_1$ & $c_2$ & $c_3$ & & & \\
 & & $-2.66^{+0.07}_{-0.08}$  & $-1.03^{+0.08}_{-0.07}$ & $-0.43^{+0.01}_{-0.01}$ & $-0.30^{+0.01}_{-0.01}$ & $-0.96^{+0.04}_{-0.04}$ & & & \\ 
 \cline{2-10} 
 & \textbf{7 bins} & $M_*(z_p)$ & $\log_{10}\Phi(z_p)$    & & & & & & \multirow{4}{*}{ 159.27 } \\
 & \multirow{3}{*}{$0.68-3.5$} & $-25.40^{+0.16}_{-0.16}$ & $-5.47^{+0.05}_{-0.06}$ & & & & & & \\
 &   & $a$ & $b$ & $c_1$ & $c_2$ & $c_3$ & & & \\
 & & $-2.67^{+0.07}_{-0.08}$  & $-1.04^{+0.08}_{-0.07}$ & $-0.44^{+0.01}_{-0.01}$ & $-0.30^{+0.01}_{-0.01}$ & $-0.97^{+0.04}_{-0.04}$ & & & \\ 
 \cline{2-10} 
 & \textbf{6 bins} & $M_*(z_p)$ & $\log_{10}\Phi(z_p)$    & & & & & & \multirow{4}{*}{ 130.82 }\\
 & \multirow{3}{*}{$0.68-3.0$} & $-25.61^{+0.17}_{-0.17}$ & $-5.54^{+0.06}_{-0.06}$ & & & & & & \\
 &           & $a$ & $b$ & $c_1$ & $c_2$ & $c_3$ & & & \\
 &            & $-2.74^{+0.08}_{-0.09}$  & $-1.10^{+0.07}_{-0.07}$ & $-0.49^{+0.02}_{-0.02}$ & $-0.32^{+0.01}_{-0.01}$ & $-1.08^{+0.04}_{-0.04}$ &                        &                           &                               \\ \hline
\multirow{18}{*}{\textbf{LEDE}$_\mathbf{7+2}$} & \textbf{8 bins} & $M_*(z_p)$ & $\log_{10}\Phi(z_p)$ & & & & & & \multirow{6}{*}{144.39}\\
     & \multirow{5}{*}{$0.68-4.0$} & $-26.17^{+0.16}_{-0.16}$ & $-5.77^{+0.06}_{-0.07}$ & & & & & & \\
     & & $c_1$ & $c_2$ & $c_3$ & & & & & \\
     & & $-0.94^{+0.06}_{-0.06}$  & $-0.4^{+0.02}_{-0.02}$  & $-1.99^{+0.1}_{-0.1}$   & & & & & \\
     & & $a(z_p)$ & $c_a$ & $b(z_p)$ & $c_b$ & & & & \\
     & & $-2.84^{+0.08}_{-0.09}$  & $-0.16^{+0.07}_{-0.07}$ & $-1.33^{+0.05}_{-0.05}$ & $-0.5^{+0.04}_{-0.04}$  & & & & \\     \cline{2-10} 
     & \textbf{7 bins} & $M_*(z_p)$ & $\log_{10}\Phi(z_p)$    & & & & & & \multirow{6}{*}{ 137.44 }\\
     & \multirow{5}{*}{$0.68-3.5$} & $-26.16^{+0.16}_{-0.16}$ & $-5.76^{+0.07}_{-0.07}$ & & & & & & \\
     & & $c_1$ & $c_2$ & $c_3$ & & & & & \\
     & & $-0.92^{+0.06}_{-0.07}$  & $-0.39^{+0.02}_{-0.02}$ & $-1.95^{+0.11}_{-0.11}$ & & & & & \\
     & & $a(z_p)$ & $c_a$ & $b(z_p)$ & $c_b$ & & & & \\
     & & $-2.84^{+0.08}_{-0.09}$  & $-0.14^{+0.07}_{-0.07}$ & $-1.33^{+0.05}_{-0.05}$ & $-0.47^{+0.05}_{-0.05}$ & & & & \\     \cline{2-10} 
     & \textbf{6 bins} & $M_*(z_p)$ & $\log_{10}\Phi(z_p)$    & & & & & & \multirow{6}{*}{119.68}\\
     & \multirow{5}{*}{$0.68-3.0$} & $-26.18^{+0.18}_{-0.17}$ & $-5.77^{+0.07}_{-0.07}$ & & & & & & \\
     & & $c_1$ & $c_2$ & $c_3$ & & & & & \\
     & & $-0.87^{+0.08}_{-0.08}$  & $-0.38^{+0.02}_{-0.02}$ & $-1.84^{+0.14}_{-0.13}$ & & & & & \\
     & & $a(z_p)$ & $c_a$ & $b(z_p)$ & $c_b$ & & & & \\
     & & $-2.85^{+0.09}_{-0.10}$  & $-0.09^{+0.06}_{-0.06}$ & $-1.33^{+0.06}_{-0.05}$ & $-0.4^{+0.06}_{-0.06}$  & & & & \\ \hline
\multirow{12}{*}{\textbf{LEDE}$_\mathbf{8}$} & \textbf{8 bins} & $M_*(z_p)$ & $\log_{10}\Phi(z_p)$    & & & & & & \multirow{4}{*}{ 141.64 }\\
     & \multirow{3}{*}{$0.68-4.0$} & $-25.72^{+0.16}_{-0.16}$ & $-5.56^{+0.06}_{-0.06}$ & & & & & & \\
     & & $a$ & $b$ & $c_1$ & $c_2$ & $c_3$ & $c_4$ & & \\
     & & $-2.79^{+0.08}_{-0.09}$  & $-1.10^{+0.07}_{-0.06}$ & $-0.37^{+0.01}_{-0.01}$ & $-0.20^{+0.01}_{-0.01}$ & $-0.78^{+0.04}_{-0.04}$ & $0.34^{+0.04}_{-0.04}$ & & \\ 
     \cline{2-10} 
     & \textbf{7 bins} & $M_*(z_p)$ & $\log_{10}\Phi(z_p)$ & & & & & & \multirow{4}{*}{ 133.70 } \\
     & \multirow{3}{*}{$0.68-3.5$} & $-25.72^{+0.16}_{-0.16}$ & $-5.56^{+0.06}_{-0.06}$ & & & & & & \\
     & & $a$ & $b$ & $c_1$ & $c_2$ & $c_3$ & $c_4$ & & \\
     & & $-2.79^{+0.08}_{-0.09}$  & $-1.10^{+0.07}_{-0.06}$ & $-0.37^{+0.02}_{-0.02}$ & $-0.20^{+0.02}_{-0.02}$  & $-0.75^{+0.04}_{-0.05}$ & $0.38^{+0.05}_{-0.05}$ & & \\ 
     \cline{2-10} 
     & \textbf{6 bins} & $M_*(z_p)$ & $\log_{10}\Phi(z_p)$    & & & & & & \multirow{4}{*}{ 118.76 }\\
     & \multirow{3}{*}{$0.68-3.0$} & $-25.75^{+0.17}_{-0.16}$ & $-5.58^{+0.06}_{-0.06}$ & & & & & & \\
     & & $a$ & $b$ & $c_1$ & $c_2$ & $c_3$ & $c_4$ & & \\
     & & $-2.80^{+0.08}_{-0.09}$  & $-1.12^{+0.07}_{-0.06}$ & $-0.40^{+0.02}_{-0.02}$ & $-0.22^{+0.02}_{-0.02}$ &     $-0.75^{+0.07}_{-0.07}$ & $0.38^{+0.07}_{-0.07}$ & & \\ \hline
\multirow{18}{*}{\textbf{LEDE}$_\mathbf{8+2}$}   & \textbf{8 bins} & $M_*(z_p)$ & $\log_{10}\Phi(z_p)$ & & & & & & \multirow{6}{*}{ 147.19 }\\
     & \multirow{5}{*}{$0.68-4.0$} & $-25.92^{+0.20}_{-0.20}$ & $-5.65^{+0.08}_{-0.09}$ & & & & & & \\
     & & $c_1$ & $c_2$ & $c_3$ & $c_4$ & & & & \\
     & & $-0.50^{+0.09}_{-0.19}$  & $-0.21^{+0.02}_{-0.06}$ & $-1.04^{+0.22}_{-0.42}$ & $0.32^{+0.06}_{-0.11}$  & & & & \\
     & & $a(z_p)$ & $c_a$ & $b(z_p)$ & $c_b$ & & & & \\
     & & $-2.83^{+0.09}_{-0.10}$  & $-0.01^{+0.09}_{-0.11}$ & $-1.22^{+0.08}_{-0.08}$ & $-0.17^{+0.09}_{-0.17}$ & & & & \\ \cline{2-10} 
     & \textbf{7 bins} & $M_*(z_p)$ & $\log_{10}\Phi(z_p)$    & & & & & & \multirow{6}{*}{ 138.06 }\\
     & \multirow{5}{*}{$0.68-3.5$} & $-25.90^{+0.20}_{-0.20}$ & $-5.64^{+0.08}_{-0.08}$ & & & & & & \\
     & & $c_1$ & $c_2$ & $c_3$ & $c_4$ & & & & \\
     & & $-0.48^{+0.09}_{-0.14}$  & $-0.21^{+0.02}_{-0.04}$ & $-0.99^{+0.23}_{-0.30}$ & $0.35^{+0.06}_{-0.08}$  & & & & \\
     & & $a(z_p)$ & $c_a$ & $b(z_p)$ & $c_b$ & & & & \\
     & & $-2.81^{+0.09}_{-0.10}$  & $0.02^{+0.10}_{-0.10}$  & $-1.21^{+0.08}_{-0.08}$ & $-0.17^{+0.09}_{-0.12}$ & & & & \\ \cline{2-10} 
     & \textbf{6 bins} & $M_*(z_p)$ & $\log_{10}\Phi(z_p)$    & & & & & & \multirow{6}{*}{ 118.97 }\\
     & \multirow{5}{*}{$0.68-3.0$} & $-25.95^{+0.23}_{-0.21}$ & $-5.67^{+0.09}_{-0.09}$ & & & & & & \\
     & & $c_1$ & $c_2$ & $c_3$ & $c_4$ & & & & \\
     & & $-0.55^{+0.11}_{-0.12}$  & $-0.24^{+0.03}_{-0.04}$ & $-1.06^{+0.29}_{-0.29}$ & $0.33^{+0.09}_{-0.09}$  & & & & \\
     & & $a(z_p)$ & $c_a$ & $b(z_p)$ & $c_b$ & & & & \\
     & & $-2.80^{+0.11}_{-0.12}$  & $0.05^{+0.11}_{-0.12}$  & $-1.24^{+0.08}_{-0.07}$ & $-0.20^{+0.09}_{-0.09}$ & & & & \\ 
\hline 
\end{longtable}

\end{document}